\documentclass{comnet-mod}
\bibpunct{[}{]}{,}{n}{,}{;}

\usepackage{amsmath,amsfonts,amssymb}

\usepackage{multirow}

\usepackage{array}

\usepackage{braket}

\usepackage{siunitx}

\usepackage{xparse}
\usepackage{tikz}
\usetikzlibrary{graphs,calc}
\usepackage{pgfplots}
\pgfplotsset{compat=newest}

\usepackage{xifthen}
\usepackage{etoolbox}
\usepackage{xspace}
\definecolor{ColorA}{rgb}{0.2235, 0.4157, 0.6941}
\definecolor{ColorB}{rgb}{0.8549, 0.4863, 0.1882}
\definecolor{ColorC}{rgb}{0.2431, 0.5882, 0.3176}
\newcommand{\NameColorA}{blue\xspace}
\newcommand{\NameColorB}{orange\xspace}
\newcommand{\NameColorC}{green\xspace}

\tikzset{orig/.style={ColorA,mark size=1.5pt}}
\tikzset{cl/.style={ColorB,mark size=1.5pt}}
\tikzset{bter/.style={ColorC,mark size=1.5pt}}

\tikzset{modea/.style={mark=*}}
\tikzset{modeb/.style={dashed,mark=square*}}

\DeclareDocumentCommand \DegDistSubFig {O{ddbm} t* m m m m D<>{2.5in} D(){dd} D''{Mode \mode: \graphname~(\modeaname/\modebname)}}{%
\def\btype{#1}%
\def\graphname{#3}%
\def\modeaname{#4}%
\def\modebname{#5}%
\def\mode{#6}%
\def\figsize{#7}%
\def\labelname{#8}%
\def\ddcaption{#9}%
\typeout{Creating figure \labelname-\graphname-\mode}%
\subfloat[\ddcaption]{%
  \IfBooleanTF{#3}{\label{fig:dd-\graphname-\mode-bter}}{\label{fig:\labelname-\graphname-\mode}}%
    \begin{tikzpicture}
      \begin{loglogaxis}[
        width=\figsize, height={\figsize-0.5in},
        label style={font=\footnotesize},
        xlabel=Degree,
        ylabel=Count (Mean per Bin),
        legend entries={Original, CL, BTER},
        legend style={font=\footnotesize},
        legend pos=south west,
        tick label style={font=\tiny},
        xminorticks=true,
        yminorticks=true,
        ]
        \addplot[orig,modea] table[x=Degree,y=Count]{data/\graphname-Orig-\btype-\mode.dat};
        \addplot[cl,modea] table[x=Degree,y=Count]{data/\graphname-CL-\btype-\mode.dat};
        \IfBooleanT{#2}{
          \addplot[bter,modea] table[x=Degree,y=Count]{data/\graphname-BTER-\btype-\mode.dat};
        }
      \end{loglogaxis}
    \end{tikzpicture}
  }
}

\DeclareDocumentCommand \MetaSubFig {O{meta} t* m m m m D(){meta} D''{Mode \mode: \graphname~(\modeaname/\modebname)}}{%
\def\btype{#1}%
\def\graphname{#3}%
\def\modeaname{#4}%
\def\modebname{#5}%
\def\mode{#6}%
\def\labelname{#7}%
\def\metacaption{#8}%
\subfloat[\metacaption]{%
  \label{fig:\labelname-\graphname-\mode}%
  \begin{tikzpicture}
    \begin{semilogxaxis}[
      width=2.5in, height=2.0in,
      label style={font=\footnotesize},
      xlabel=Degree,
      ylabel=Metamorphosis (Mean per Bin),
      legend entries={Original, CL, BTER},
      legend style={font=\footnotesize},
      tick label style={font=\tiny},
      scaled ticks={false},
      y tick label style={/pgf/number format/.cd,fixed,fixed zerofill,precision=2}
      ]
      \addplot[modea,orig] table[x=Degree,y=Meta]{data/\graphname-Orig-\btype-\mode.dat};
      \addplot[modea,cl] table[x=Degree,y=Meta]{data/\graphname-CL-\btype-\mode.dat};
      \IfBooleanT{#2}{
        \addplot[modea,bter] table[x=Degree,y=Meta]{data/\graphname-BTER-\btype-\mode.dat};
      }
    \end{semilogxaxis}
  \end{tikzpicture}
}
}

\tikzset{vtx/.style={circle, draw=black, inner sep = 1 pt, font=\footnotesize, minimum size = 8 pt}}
\tikzset{ltvtx/.style={fill=ColorA!50}}
\tikzset{rtvtx/.style={fill=ColorB!50}}

\DeclareDocumentCommand \MakeGroup {D(){} D<>{} D<>{} m m O{2} O{1} }{
  \def\gid{#1} %
  \def\lttxt{#2} %
  \def\rttxt{#3} %
  \def\nlt{#4} %
  \def\nrt{#5} %
  \def\hdelta{#6} %
  \def\vdelta{#7} %

  \coordinate (LtOffset) at (-\hdelta/2,\nlt*\vdelta/2+\vdelta/2);
  \foreach \lt in {1,...,\nlt}
  \node[vtx,ltvtx] (\gid L\lt) at ($(0,-\lt*\vdelta)+(LtOffset)$) {\lttxt};

  \coordinate (RtOffset) at (\hdelta/2,\nrt*\vdelta/2+\vdelta/2);
  \foreach \rt in {1,...,\nrt}
  \node[vtx,rtvtx] (\gid R\rt) at ($(0,-\rt*\vdelta)+(RtOffset)$) {\rttxt};
}

\DeclareDocumentCommand \MakeEdges { D(){} m m D<>{1} O{0.5} }{
  \def\gid{#1} %
  \def\nlt{#2} %
  \def\nrt{#3} %
  \pgfmathsetseed{#4};
  \def\thresh{#5}

  \foreach \lt in {1,...,\nlt}
  \foreach \rt in {1,...,\nrt}
  {
    \pgfmathparse{rnd}
    \ifdimless{\pgfmathresult pt}{\thresh pt}{\draw (\gid L\lt) -- (\gid R\rt);}{}
  }

}
    
\usepackage{pgfplots}

\usepackage[font=footnotesize,justification=Centering,singlelinecheck=false]{subfig}

\usepackage{xcolor}

\usepackage[draft]{hyperref}
\hypersetup{
  colorlinks=true,
  allcolors=green!50!black,
  urlcolor=red!50!black
}

\usepackage[capitalize,nameinlink]{cleveref}

\crefformat{equation}{\textup{#2(#1)#3}}
\crefrangeformat{equation}{\textup{#3(#1)#4--#5(#2)#6}}
\crefmultiformat{equation}{\textup{#2(#1)#3}}{ and \textup{#2(#1)#3}}
{, \textup{#2(#1)#3}}{, and \textup{#2(#1)#3}}
\crefrangemultiformat{equation}{\textup{#3(#1)#4--#5(#2)#6}}%
{ and \textup{#3(#1)#4--#5(#2)#6}}{, \textup{#3(#1)#4--#5(#2)#6}}{, and \textup{#3(#1)#4--#5(#2)#6}}

\Crefformat{equation}{#2Equation~\textup{(#1)}#3}
\Crefrangeformat{equation}{Equations~\textup{#3(#1)#4--#5(#2)#6}}
\Crefmultiformat{equation}{Equations~\textup{#2(#1)#3}}{ and \textup{#2(#1)#3}}
{, \textup{#2(#1)#3}}{, and \textup{#2(#1)#3}}
\Crefrangemultiformat{equation}{Equations~\textup{#3(#1)#4--#5(#2)#6}}%
{ and \textup{#3(#1)#4--#5(#2)#6}}{, \textup{#3(#1)#4--#5(#2)#6}}{, and \textup{#3(#1)#4--#5(#2)#6}}

\usepackage{algorithm,algpseudocode}
\Crefname{ALC@unique}{Line}{Lines}

\usepackage{mathtools,pict2e,epic,relsize}

\newcommand*\ctplr{\DOTSB\mathbin{\mathpalette\ctplraux\relax}}
\newcommand*\btrfly{\DOTSB\mathbin{\mathpalette\btrflyaux\relax}}

\newlength\ztimespadding
\newcommand*\ltaux[2]{\vcenter{\hbox{\hspace{\ztimespadding}%
  \ifx#1\displaystyle
    \setlength\unitlength{1ex}%
    \linethickness{.1ex}%
  \else\ifx#1\textstyle
    \setlength\unitlength{1ex}%
    \linethickness{.1ex}%
  \else\ifx#1\scriptstyle
    \setlength\unitlength{.8ex}%
    \linethickness{.09ex}%
  \else\ifx#1\scriptscriptstyle
    \setlength\unitlength{.65ex}%
    \linethickness{.07ex}%
  \fi\fi\fi\fi
  \begin{picture}(1,1)\roundcap
    \put(0.5,0.5){\circle{1.5}}
    \put(0,0){\relsize{-4}{1}}
  \end{picture}%
  \hspace{\ztimespadding}%
}}}
\newcommand*\rtaux[2]{\vcenter{\hbox{\hspace{\ztimespadding}%
  \ifx#1\displaystyle
    \setlength\unitlength{1ex}%
    \linethickness{.1ex}%
  \else\ifx#1\textstyle
    \setlength\unitlength{1ex}%
    \linethickness{.1ex}%
  \else\ifx#1\scriptstyle
    \setlength\unitlength{.8ex}%
    \linethickness{.09ex}%
  \else\ifx#1\scriptscriptstyle
    \setlength\unitlength{.65ex}%
    \linethickness{.07ex}%
  \fi\fi\fi\fi
  \begin{picture}(1,1)\roundcap
    \put(0.5,0.5){\circle{1.5}}
    \put(0,0){\relsize{-4}{2}}
  \end{picture}%
  \hspace{\ztimespadding}%
}}}

\newcommand*\ctplraux[2]{\vcenter{\hbox{\hspace{\ztimespadding}%
  \ifx#1\displaystyle
    \setlength\unitlength{1ex}%
    \linethickness{.1ex}%
  \else\ifx#1\textstyle
    \setlength\unitlength{1ex}%
    \linethickness{.1ex}%
  \else\ifx#1\scriptstyle
    \setlength\unitlength{.8ex}%
    \linethickness{.09ex}%
  \else\ifx#1\scriptscriptstyle
    \setlength\unitlength{.65ex}%
    \linethickness{.07ex}%
  \fi\fi\fi\fi
  \begin{picture}(1,1)\roundcap
    \put(0,0){\line(1,0){1}}
    \put(1,0){\line(-1,1){1}}
    \put(0,1){\line(1,0){1}}
  \end{picture}%
  \hspace{\ztimespadding}%
}}}

\newcommand*\btrflyaux[2]{\vcenter{\hbox{\hspace{\ztimespadding}%
  \ifx#1\displaystyle
    \setlength\unitlength{1ex}%
    \linethickness{.1ex}%
  \else\ifx#1\textstyle
    \setlength\unitlength{1ex}%
    \linethickness{.1ex}%
  \else\ifx#1\scriptstyle
    \setlength\unitlength{.8ex}%
    \linethickness{.09ex}%
  \else\ifx#1\scriptscriptstyle
    \setlength\unitlength{.65ex}%
    \linethickness{.07ex}%
  \fi\fi\fi\fi
  \begin{picture}(1,1)\roundcap
    \put(0,0){\line(1,0){1}}
    \put(1,0){\line(-1,1){1}}
    \put(0,1){\line(1,0){1}}
    \put(0,0){\line(1,1){1}}
  \end{picture}%
  \hspace{\ztimespadding}%
}}}
\newcommand{\qtext}[1]{\ensuremath\quad\text{#1}\quad}
\newcommand{\ntri}{n^{\triangle}}
\newcommand{\nwdg}{n^{\wedge}}
\newcommand{\nbtf}{n^{\btrfly}}
\newcommand{\ncat}{n^{\ctplr}}

\newcommand{\lt}{u}
\newcommand{\rt}{v}

\begin{document}

\title{Measuring and Modeling Bipartite Graphs with Community Structure%
}

\shorttitle{Measuring and Modeling Bipartite Graphs}
\shortauthorlist{S. Aksoy, T. G. Kolda, and A. Pinar}

\author{%
  \name{Sinan Aksoy}
  \address{Department of Mathematics, University of California, San Diego, CA}
  \name{Tamara G. Kolda}
  \address{Sandia National Laboratories, Livermore, CA. Corresponding author: tgkolda@sandia.gov}
  \and
  \name{Ali Pinar}
  \address{Sandia National Laboratories, Livermore, CA}
}
\maketitle

\begin{abstract}
{Network science is a powerful tool for analyzing complex systems 
in fields ranging from sociology to engineering to biology.
This paper is focused on generative models of large-scale \emph{bipartite
  graphs}, also known as two-way graphs or two-mode networks. %
We propose two generative models that can be easily tuned to reproduce
the characteristics of real-world networks, not just qualitatively,
but quantitatively.
The characteristics we consider are the degree distributions and the metamorphosis coefficient.
The metamorphosis coefficient, a bipartite analogue of the clustering coefficient,
is the proportion of length-three paths that participate in length-four cycles.
Having a high metamorphosis coefficient is a necessary condition for close-knit community structure.
We define edge, node, and degreewise metamorphosis coefficients, enabling a
more detailed understanding of the bipartite connectivity that is not explained by degree distribution alone.
Our first model, bipartite Chung-Lu (CL), is able to reproduce real-world degree
distributions, and our second model,  bipartite block two-level Erd\"os-R\'enyi (BTER), 
reproduces both the degree distributions as well as the degreewise metamorphosis coefficients.
We demonstrate the effectiveness of these models on several real-world data sets.}
{bipartite generative graph model, two-way graph model, two-mode network, %
 metamorphosis coefficient, bipartite clustering coefficient, complex networks}
\end{abstract}

\section{Introduction}
\label{sec:introduction}
Network science is a powerful tool for analyzing complex systems in fields ranging from sociology to engineering to biology. 
The ability to develop realistic models of the networks is needed for several reasons. 
Pragmatically, we need to generate artificial networks to facilitate sharing of realistic network data while respecting concerns about privacy and security of data. 
More generally, generative models enable unlimited network data generation for computational analysis, e.g., varying the characteristics of the graph to test a graph algorithm under different scenarios. 
Ultimately, we hope to \emph{understand} the underlying nature of complex systems, and modeling them mathematically is a way to test our understanding.

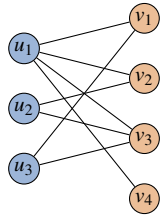
\begin{figure}[htp]
  \centering
\begin{tikzpicture}[scale=0.8]
  \MakeGroup<$u_{\lt}$><$v_{\rt}$>{3}{4}
  \MakeEdges{3}{4}<2>[.5]
\end{tikzpicture}
%
%
%
%
  \caption{Bipartite graph}
  \label{fig:bipartite}
\end{figure}

This paper is focused on generative models of \emph{bipartite
  graphs}, also known as \emph{two-way graphs} or \emph{two-mode networks}. %
Many real-world systems are naturally expressed as bipartite graphs. 
The defining characteristic of a bipartite graph is that its vertices are divided into two
partitions, $U$ and $V$, such that edges, $E$, only connect vertices across the two
partitions, i.e.,
\begin{displaymath}
  G = (U, V, E) 
  \qtext{with}
  U \cap V = \emptyset 
  \qtext{and}
  E \subseteq  U \otimes V.
\end{displaymath}
Examples of bipartite graphs include author-paper networks,
user-product purchase histories, user-song play lists, actor-movie
connections, document-keyword mappings, and so on.  
Hypergraphs can be represented as bipartite graphs in the sense of an \emph{incidence graph}: the nodes and hyperedges are represented by $U$ and $V$, respectively, and edge $(i,j)$ exists if node $i$ is in hyperedge $j$.
Bipartite graphs
have been widely studied; see
\cite{Dh01,GuLa04,RoAl04,SuQuChFa05,GuLa06,LaMaVe08} and references therein.
An example bipartite graph is shown in \cref{fig:bipartite}.

\begin{figure}[htp]
  \centering
  \subfloat[Caterpillars]{
    \begin{tikzpicture}
      \MakeGroup{2}{2}[1]
      \draw (L1)--(R1)--(L2)--(R2);
    \end{tikzpicture}
    ~~~~~
    \begin{tikzpicture}
      \MakeGroup{2}{2}[1]
      \draw (R1)--(L1)--(R2)--(L2);
    \end{tikzpicture}
    ~~~~~
    \begin{tikzpicture}
      \MakeGroup{2}{2}[1]
      \draw (L2)--(R1)--(L1)--(R2);
    \end{tikzpicture}
    ~~~~~
    \begin{tikzpicture}
      \MakeGroup{2}{2}[1]
      \draw (L1)--(R2)--(L2)--(R1);
    \end{tikzpicture}
  }
  ~~~~~~~~~~
  \subfloat[Butterfly\label{fig:butterfly}]{
    ~~~
    \begin{tikzpicture}
      \MakeGroup{2}{2}[1]
      \draw (L1)--(R2)--(L2)--(R1)--(L1);
    \end{tikzpicture}
    ~~~~~
  }
  \caption{A butterfly contains four distinct caterpillars}
  \label{fig:structures}
\end{figure}
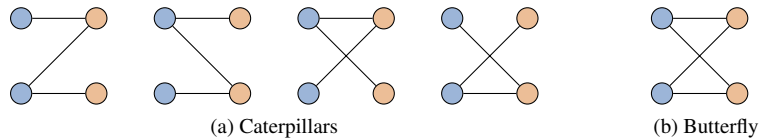

We propose a generative model that can be easily tuned to reproduce
the characteristics of real-world networks, not just qualitatively,
but quantitatively.
The measurements we consider in this paper are the degree distributions and the bipartite analog of the
clustering coefficient. 
Of course, there are many other measurements that we could consider, and we come back to 
those shortly.
The degree distribution is the number (or proportion) of nodes of degree $d$ for each $d=1,2,\dots$.
A bipartite graph has \emph{two} degree distributions, one for each vertex partition. 
As we see in the results in \cref{sec:experimental-results-1}, 
these two distributions may be quite different from one another, 
in part because the size (and, consequently, the average degree) 
for each partition may be quite different.
The clustering coefficient of a one-way graph, introduced by Holland and Leinhardt \cite{HoLe70},
is the probability that a two-path (or \emph{wedge}) participates in a
three-cycle (or \emph{triangle}); i.e.,
\begin{displaymath}
  c 
  = \frac{3\ntri}{\nwdg}
  = \frac{3 \times ( \text{total number of triangles} )}{\text{total number of wedges}}.
\end{displaymath}
One characteristic of a bipartite graph is that is has no odd-length cycles; hence, it cannot have a triangle.
Robins and Alexander \cite{RoAl04} propose a bipartite clustering coefficient
that is the
probability that a bipartite three-path (or \emph{caterpillar}%
) 
participates in a bipartite four-cycle (or \emph{butterfly}%
), i.e.,
\begin{equation}\label{eq:meta}
  c
  = \frac{4\nbtf}{\ncat} 
  = \frac{4 \times ( \text{total number of butterflies} )}{\text{total number of caterpillars}}.
\end{equation}
We call $c$ in \cref{eq:meta} the \emph{metamorphosis coefficient} in reference to the notion of how often caterpillars become butterflies. 
Opsahl~\cite{Op13} says ``it could be considered a measure of reinforcement between two individuals rather than clustering of a group of individuals.''
The multiplier of four in the numerator is because every butterfly contains four
distinct caterpillars, just as every triangle contains three distinct
wedges; see \cref{fig:structures}.  
A high metamorphosis coefficient in a bipartite graph is
indicative of greater community structure in the graph, analogous to
the role of a high clustering coefficient in a one-way graph.
The metamorphosis coefficient plays an important role in communities because any tight-knit community will have a high metamorphosis coefficient.
In \cref{sec:degr-metam-coeff}, we extend \cref{eq:meta} to define edge, node, and degreewise metamorphosis coefficients, enabling a
more detailed understanding of the bipartite community structure.

As mentioned above, there are numerous more complex metrics that can
be used in the analysis of bipartite graphs; see, e.g.,
\cite{RoAl04,LaMaVe08}.
Examples of such metrics include graph diameter (i.e., the maximum
distance between any pair of nodes), singular values of the adjacency
matrix, 
centrality measures, joint degree distributions (i.e., the proportion of nodes of
degree $d$ that connect to nodes of degree $d'$ for all possible pairs
$(d,d')$), assortativity (i.e., degree correlations), subgraph census 
(i.e., counts of all patterns of three nodes, four nodes, and so on).
Moreover, we can consider the analysis of the \emph{product graphs}
that are produced by considering just \emph{one} vertex partition
and connecting any two nodes that share a common neighbor in the
bipartite graph.
Because these quantities are often expensive to compute for large
graphs, we do not consider them in this paper.
Instead, we consider the simpler metrics of degree distribution and metamorphosis coefficients here for reasons
analogous to those considered by \cite{LaMaVe08,SeKoPi12}.  The degree
distributions are interesting because many large real-world graphs
(including the ones we consider here) do not have the Poisson
distributions that occur if edges are distributed at random; instead,
they have heavy-tailed distributions (hypothesized to be power law
\cite{BaAl99} or log-normal \cite{KoPiPlSe14}). The metamorphosis
coefficient is interesting because random graphs that have the same
degree distribution as a real-world graph tend to have a much lower
metamorphosis coefficients, as we show in
\cref{sec:shortcomings-cl}. This is a well-known phenomena for the
clustering coefficient of one-way graphs \cite{Ne03}. We contend that
matching metamorphosis coefficients is critical for capturing social
or, more generally, interconnectedness behavior in networks.

With the goals of capturing both degree distribution and our
newly-defined degreewise metamorphosis coefficients, we develop two
different models.
The first is a straightforward
extension of Chung-Lu (CL) \cite{AiChLu01,ChLu02,ChLu02b}, 
which is very closely related to
the configuration model.
Our experimental results show that 
this model is
effective at reproducing the degree distributions.
However, the bipartite CL graphs do
not produce the same metamorphosis coefficients as observed in
real-world networks.  Therefore, the second model we propose is a
bipartite extension of the Block Two-Level Erd\H{o}s-R\'{e}nyi (BTER)
\cite{SeKoPi12,KoPiPlSe14} model.  The BTER model is a good starting
point because it reproduces both the degree distribution and
degreewise clustering coefficients of a given network. To do so, it
groups nodes into Erd\H{o}s-R\'{e}nyi (ER) subgraphs, called affinity
blocks, that are highly connected and so produce high clustering
coefficients.  We propose a bipartite BTER that reproduces
both the degree distributions and the degreewise metamorphosis
coefficients. This extension is not straightforward since the affinity 
block concept does not carry over easily to the bipartite case, so we 
develop a new method for creating the blocks. Computational results for bipartite BTER show that it achieves our goals of matching both the degree distributions and the degreewise metamorphosis coefficients.

Before we continue, we briefly survey related work.
As mentioned above, the proposed bipartite CL model is very similar to 
bipartite configuration models.
Both Newman, Strogatz, and Watts \cite{NeStWa01} and
Guillaume and Latapy \cite{GuLa06} propose bipartite configuration models
that \emph{sample} node degrees from a distribution (one per partition), create stubs for each node equal to its degree, and then match the stubs.
Our bipartite CL model is very similar, and we discuss the connections 
further in \cref{sec:chung-lu-two}.
On the theoretical side,
Kannan, Tetali, and Vempala \cite{KaTeVe99} consider the convergence 
of a Markov 
chain rewiring algorithm for generating a bipartite graph 
and show that their process is rapidly mixing for regular bipartite graphs, i.e., graphs where all the node degrees are the same.
Later, this result was extended to graphs in which only one partition 
is required to be regular~\cite{MiErSo13}. 
In statistics, researchers consider the problem of generating
\emph{binary contingency tables} with given row and columns sums. 
This problem  is equivalent to specifying the degrees in a bipartite graph.
In \cite{BeBhVi06, ChDiHoLi05}, the focus is on an algorithm that is \emph{guaranteed} to produce a specified row and column sums, assuming 
its realizable. The proposed algorithm is not considered practical but 
rather of theoretical interest.
As for one-way graphs, there are also \emph{incremental growth} models
in which the output graph is iteratively constructed by adding
vertices and edges according to some rule. 
The methods are sequential since the connections created for new nodes
usually depend on the state of the graph at the iteration in which the 
nodes are created.
Guillaume and Latapy \cite{GuLa06}
analyze a bipartite growing model in which new vertices are linked via a 
preferential attachment process. 
Other work has considered a similar idea except that 
one vertex partition remains fixed while the other grows~\cite{PeChoMuGa07,GaGhKrSr12}.
Lastly, we note a variety of application-specific generative bipartite graph models 
have been introduced to model specific networks, 
including pollination
networks in ecology \cite{SaReUz08, DoFrBlGr09}, and protein-domain networks in 
biology \cite{NaOcHaAk09}.

\section{Data sets}
\label{sec:data-sets}

We test our methods on publicly-available real-world datasets, whose
properties are summarized in \cref{tab:graphs}. Their degree
distributions are shown in \cref{fig:dd1,fig:dd2,fig:dd3}.
\textbf{CondMat} represents an author-paper network from arXiv preprints in
condensed matter physics from 1995--99 \cite{Ne01}; this has mostly
been used in the context of the coauthorship graph, but here we
consider the underlying data \cite{ToreOpsahl}.  The majority of
authors have only a single paper, whereas the most prolific author has
116 papers. Conversely, the most coauthors on a
single paper is 18, and the most likely scenario is for a paper to
have 2 or 3 authors.
\textbf{IMDB} links movies and the actors that appeared in them
\cite{LaMaVe08,La06}, as collected from the Internet Movie Database.
The busiest actor was in 294 movies; conversely, the largest
production had 646 actors. 
\textbf{Flickr} \cite{MiMaGuDr07,Mi15-PC} is an online photo-sharing site, and
the network represents group membership of various users.  The most connected
user is in 2186 groups, whereas the largest group has 34989
members.
\textbf{MovieLens} \cite{HaKo15,movielens} is a very famous dataset that links
movies and their reviewers/critics. The most-reviewed movie had 34864
reviews, and the most active critic reviewed 7359 movies
(\NameColorB). This dataset apparently excludes critics with less than 20 reviews.
The \textbf{MillionSong} \cite{BeElWhLa11,McBeElLa12,echonest} dataset connects
users and the songs they played. The dataset only includes listeners
of ten songs or more. The widest-ranging user listened to 4,400 distinct
songs. The top song was played by 110,479 distinct listeners.
The \textbf{Peer2Peer} dataset \cite{LaMaVe08,La06} links users (peers) and the
files they uploaded or downloaded. The busiest user touched 19,496
files. On the other hand, the most popular file only had 3396
downloads.
\textbf{LiveJournal} \cite{MiMaGuDr07,Mi15-PC} represents user-group
memberships from a blogging site.  The most engaged user is in 300
groups, which appears to be the maximum allowed, since there are five
persons in that category.  The largest group has over one million members.

\sisetup{
  input-decimal-markers = .,input-ignore = {,},table-number-alignment = right,
  group-separator={,}, group-four-digits = true
}
\begin{table}[th]
\setlength\extrarowheight{2pt}
  \centering\footnotesize
  \caption{Real-world bipartite graphs}
  \label{tab:graphs}
  \begin{tabular}{|l|*{2}{S[table-format = 7.0]@{~}l|}S[table-format = 9.0]|}
    \hline
    \multicolumn{1}{|c|}{\bf Name} & 
    \multicolumn{2}{c|}{\bf Partition 1} & 
    \multicolumn{2}{c|}{\bf Partition 2} & 
    \multicolumn{1}{c|}{\bf Edges} \\ \hline
    CondMat \cite{Ne01,ToreOpsahl} & 16726 & authors & 22016 & papers & 58595  \\ \hline
    IMDB \cite{LaMaVe08,La06}& 127823 & actors & 383640 & movies & 1470418  \\ \hline
    Flickr \cite{MiMaGuDr07,Mi15-PC} & 1728701 & users & 103648 & groups & 8545307 \\ \hline
    MovieLens \cite{HaKo15,movielens} & 65133 & movies & 71567 &critics & 10000054 \\ \hline
    MillionSong \cite{BeElWhLa11,McBeElLa12,echonest} & 1019318 & users & 384546 & songs & 48373586  \\ \hline
    Peer2Peer \cite{LaMaVe08,La06} & 1986588 & peers & 5380546 & files & 55829392 \\ \hline
    LiveJournal \cite{MiMaGuDr07,Mi15-PC} & 5284451 & users & 7489296 & groups & 112307385 \\ \hline
  \end{tabular}
\end{table}

\section{Notation}
\label{sec:notation}

We set up the notation for one-way and two-way graphs. 
We assume  all graphs are {\em simple}, meaning that there are no
multiple edges.
In the one-way case, we let $n=|V|$ and index nodes in $V$ by $i,j\in \set{1,\dots,n}$. 
In the two-way case, we let $n^{\lt} = |U|$ and $n^{\rt}=|V|$ denote the sizes of partitions one (left) and two (right), respectively. 
We use $i\in \set{1,\dots,n^{\lt}}$ and $j\in \set{1,\dots,n^{\rt}}$ to index nodes in partition one and two, respectively. 
Without loss of generality, indexing by $i$ assumes partition one and likewise for $j$ and partition two.
For instance, if $i=j=2$ in the two-way case, although vertices $i$ and $j$ have the same vertex label of $2$, they belong to different partitions
and are thus two distinct vertices.

\begin{table}[htbp]
  \setlength\extrarowheight{2pt}
 \caption{Notation}
  \centering
  \begin{tabular}{|r@{~:~}l|r@{~:~}l|}
    \hline
    \multicolumn{2}{|c|}{\bf One-Way Graph} &
    \multicolumn{2}{c|}{\bf Two-Way Graph} \\ 
    \multicolumn{2}{|c|}{$G=(V,E)$} &
    \multicolumn{2}{c|}{$G=(U,V,E)$} \\ [2mm]
    Vertices & $V$ & Vtx.~Partition 1 & $U$ \\
    \multicolumn{2}{|c|}{} & Vtx.~Partition 2 & $V$  \\ 
    Edges & $E \subseteq V \otimes V$ & Edges & $E \subseteq U \otimes V$  \\ 
    \# Vertices & $n=|V|$ & \# Vertices in $U$ & $n^{\lt} = |U|$ \\ 
    \multicolumn{2}{|c|}{} &\# Vertices in $V$ & $n^{\rt} = |V|$ \\ 
    \# Edges & $m=|E|$ &    \# Edges & $m=|E|$ \\
    \# Wedges & $\nwdg$ & \# Caterpillars & $\ncat$ \\
    \# Triangles & $\ntri$ & \# Butterflies & $\nbtf$ \\ 
    Clust.~Coeff. & $c = 3\ntri/\nwdg$ & Meta.~Coeff. & $c = 4\nbtf/\ncat$ \\
    Vertex Index & $i \in \set{1,\dots,n}$ & Index in $U$ & $i \in \set{1,\dots,n^{\lt}}$ \\
    \multicolumn{2}{|c|}{} & Index in $V$ & $j \in \set{1,\dots,n^{\rt}}$ \\
    Degree of $i$ & $d_i = | \set{j \in V | (i,j) \in E} |$ &  Degree of $i$ & $d_i^{\lt} = | \set{j \in V | (i,j) \in E} |$ \\
    \multicolumn{2}{|c|}{} & Degree of $j$ & $d_j^{\rt} = | \set{i \in U | (i,j) \in E} |$ \\
    \hline
  \end{tabular}
  \label{tab:notation}
\end{table}

\section{Fast Bipartite Chung-Lu Model}
\label{sec:fast-bipartite-chung}

We adapt the Chung-Lu generative model~\cite{AiChLu01,ChLu02,ChLu02b} to bipartite graphs and demonstrate its ability to reproduce bipartite degree distributions.
We follow the notation described in \cref{sec:notation}.

\subsection{Chung-Lu for One-way Graphs}
\label{sec:chung-lu-one}

Consider a one-way graph $G=(V,E)$.  The Chung-Lu (CL) model
attempts to match the desired degrees
$\set{d_1,\dots,d_n}$ where $d_i$ denotes the desired degree of vertex
$i$. The model generates a random graph on $n$ vertices such that the
probability that vertex $i$ is adjacent to $j$ is given by
\begin{displaymath}
  \Pr((i, j)\in E)=\frac{d_id_j}{2m} 
  \qtext{where} 
  m = \frac{1}{2}\sum_{i=1}^n d_i = \text{ desired number of edges}.
\end{displaymath}
To ensure the quantities are all true probabilities, we assume $d_i \leq \sqrt{2m}$ for all $i$.
A classical implementation of the CL model on $n$ vertices flips a
coin for each of the ${n \choose 2}=\Omega(n^2)$ possible edges. Many
real-world graphs are large and sparse, i.e., the number of edges
$m=O(n)$. For this reason, we favor ``fast'' CL that flips only $2m$ coins
\cite{DuKoPiSe13,KoPiPlSe14}.  
We explain the fast method below in the context of two-way graphs.

\subsection{Chung-Lu for Two-way Graphs}
\label{sec:chung-lu-two}

Consider the bipartite graph $G=(U,V,E)$. Here we have separate
desired degrees for the vertices in $U$ and $V$, denoted
\begin{displaymath}
  \set{d_i^{\lt}}_{i=1}^{n^{\lt}}
  \qtext{and} 
  \set{d_j^{\rt}}_{j=1}^{n^{\rt}},
\end{displaymath}
respectively.
Necessarily, the sums of the degrees in each partition must be equal to each other and to the number of edges, i.e.,
\begin{displaymath}
  m = \sum_{i=1}^{n^{\lt}} d_i^{\lt} = \sum_{j=1}^{n^{\rt}} d_j^{\rt}.
\end{displaymath}
Hence, the bipartite CL model generates a random bipartite graph on $n^{\lt}$ vertices in the first partition and $n^{\rt}$ vertices in the second partition such that 
\begin{equation}\label{eq:bcl}
  \Pr((i,j) \in E) = \frac{d_i^{\lt} d_j^{\rt}}{m}.
\end{equation}
To ensure these are true probabilities, we assume $d_i^{\lt} \leq \sqrt{m}$ for all $i$ and 
$d_j^{\rt} \leq \sqrt{m}$ for all $j$.
A na\"ive implementation of bipartite Chung-Lu would flip a coin for all $n^{\lt}n^{\rt}$ possible edges. Instead, we adopt the ``fast'' approach as follows. Rather than flipping a coin for every possible edge, we instead randomly choose two endpoints for every expected edge. Since the graph is sparse, we may assume that $m \ll n^{\lt}n^{\rt}$, so this approach requires many fewer random samples. For each of the $m$ edges, we choose endpoints in $U$ and $V$ proportional to
\begin{displaymath}
  \Pr(i) = \frac{d_i^{\lt}}{m}
  \qtext{and}
  \Pr(j) = \frac{d_j^{\rt}}{m},
\end{displaymath}
respectively. The probability that edge $(i,j)$ exists is then given by
\begin{displaymath}
  \Pr((i,j)\in E) = m \cdot \Pr(i) \cdot \Pr(j) = \frac{d_i^{\lt} d_j^{\rt}}{m},
\end{displaymath}
which is the same as \cref{eq:bcl}.
Although both implementations of CL yield identical expected degrees, a key distinction is that multiple edges between the same pair of vertices are possible in fast CL.
In practice, for large graphs with heavy-tailed degree distributions,
this rarely presents a problem.
The fast bipartite CL algorithm is described in \cref{alg:fbcl}.

\begin{algorithm}[htbp]
  \caption{Fast Bipartite CL}
  \label{alg:fbcl}
  \begin{algorithmic}[1]
    \Procedure{fbcl}{$\set{ d^{\lt}_i }$, $\set{ d_j^{\rt} }$}
    \State $m \gets \sum_i d_i^{\lt}$
    \State $E \gets \emptyset$
    \For{$k=1,\dots,m$}
    \State Randomly select $i \in U$ proportional to $d_i^{\lt}/m$
    \State Randomly select $j \in V$ proportional to $d_j^{\rt}/m$
    \State $E \gets E \cup (i,j)$
    \Comment Duplicate edges discarded
    \EndFor \\
    \Return $E$
    \EndProcedure
  \end{algorithmic}
\end{algorithm}

As mentioned in the introduction, a 
closely-related approach is the bipartite configuration model
\cite{NeStWa01,GuLa06}. These differ in some details; for instance, in the CL model, the
degree distribution is not specified exactly but rather each degree is sampled from a distribution. However, if we ignore that detail, we may consider that each node $i\in U$ has $d_i^{\lt}$ stubs and likewise each
node $j \in V$ has $d_j^{\rt}$ stubs, and the stubs from partition one are
randomly connected to the stubs in partition two. 
As with fast bipartite CL, we discard any repeated edges.
In all three cases, bipartite CL, fast bipartite CL, 
and the bipartite configuration models,  the expected degree of a vertex is
the same.

\subsection{Experimental Results}
\label{sec:experimental-results-1}

We generate random graphs using CL and the degree distributions of
the graphs described in \cref{sec:data-sets}.
The degree distributions are shown in \cref{fig:dd1,fig:dd2,fig:dd3}. 
\newcommand{\ddcaption}{Degree distributions illustrating the original (\NameColorA) versus bipartite fast Chung-Lu (\NameColorB). The data is log-binned.}%
\begin{figure}[th]
  \centering
  \DegDistSubFig{CondMat}{Authors}{Papers}{1}
  \DegDistSubFig{CondMat}{Authors}{Papers}{2}

  \caption{\ddcaption}
  \label{fig:dd1}
\end{figure}%
\begin{figure}[p]
  \centering
  \DegDistSubFig{IMDB}{Actors}{Movies}{1}
  \DegDistSubFig{IMDB}{Actors}{Movies}{2}

  \DegDistSubFig{Flickr}{Users}{Groups}{1}
  \DegDistSubFig{Flickr}{Users}{Groups}{2}

  \DegDistSubFig{MovieLens}{Movies}{Critics}{1}
  \DegDistSubFig{MovieLens}{Movies}{Critics}{2}

  \caption{\ddcaption}
  \label{fig:dd2}
\end{figure}%
\begin{figure}[p]
  \centering
  \DegDistSubFig{MillionSong}{Users}{Songs}{1}
  \DegDistSubFig{MillionSong}{Users}{Songs}{2}

  \DegDistSubFig{Peer2Peer}{Peers}{Files}{1}
  \DegDistSubFig{Peer2Peer}{Peers}{Files}{2}

  \DegDistSubFig{LiveJournal}{Users}{Groups}{1}
  \DegDistSubFig{LiveJournal}{Users}{Groups}{2}
  \caption{\ddcaption}
  \label{fig:dd3}
\end{figure}%
The degree distribution
of the original graph is shown in \NameColorA, and the
degree distribution of the graph generated by bipartite CL is shown in \NameColorB.
These are \emph{binned} degree distributions, as
advocated in \cite{Mi10}. We use powers of two for the bin borders, so the
$x$-coordinate $2^k$ corresponds to the bin from $[2^k,2^{k+1})$. The
$y$-coordinate is the average value for that bin, including zero
values, so the $y$-coordinate can be less than one. 
Overall, the degree distributions are very close, especially for IMDB (\cref{fig:dd-IMDB-1,fig:dd-IMDB-2}), Flickr (\cref{fig:dd-Flickr-1,fig:dd-Flickr-2}), and LiveJournal (\cref{fig:dd-LiveJournal-1,fig:dd-LiveJournal-2}).
For CondMat (\cref{fig:dd-CondMat-1,fig:dd-CondMat-2}), the distribution of degrees on the author nodes is a close match, but there is some trouble matching the paper degree distribution. The model slightly overestimates at the higher end of the degree scale and underestimates at the lower end. This is largely due to the small size of the graph and the very small distribution (maximum of 18 authors for a paper).
For MovieLens (\cref{fig:dd-MovieLens-1,fig:dd-MovieLens-2}), the model generates a few ``critic'' nodes of degree less than 20, even though no nodes exist in the true degree distribution. A similar phenomena occurs for the ``user'' nodes in MillionSong. In general, the CL model cannot handle gaps in the degree distribution because of Poisson distributions in its expectations.
In Peer2Peer (\cref{fig:dd-Peer2Peer-1,fig:dd-Peer2Peer-2}), the number of degree-1 peer nodes is underestimated, for reasons described in \cite{DuKoPiSe13}.

\subsection{Shortcomings of bipartite CL}
\label{sec:shortcomings-cl}
Overall, if we provide the degree distributions of real-world graph, the fast bipartite CL model generates a random graph whose degree distribution closely matches the degree distribution of the original graph. 
However, the graphs generated by CL typically have many fewer butterflies
than the original graphs.  
\Cref{tab:meta} shows that the number of butterflies from bipartite CL is smaller than the original graph in every case, and sometimes by one or more orders of magnitude (CondMat, IMDB, Peer2Peer).
  The number of caterpillars for the original and the generated graphs on the other hand,  is closer, so the metamorphosis of the original graph is much higher than for the generated graph with low butterfly counts. This indicates that the bipartite CL model is
omitting some important structure.

Note that the butterfly structure is what underlies the cohesive, close-knit structure in many real-world  graphs. Just as a triangle can be considered as  the smallest unit of cohesion on one-way graphs,  butterflies can be considered as  the smallest unit of cohesion in  bipartite graphs. Conversely, without butterflies, a bipartite graph will not have a community structure.  This is our motivation for considering a more complex model in the next section.

\sisetup{scientific-notation = true,group-separator={},table-format=1.2e+1,output-exponent-marker = \text{e},exponent-product={}}%
\begin{table}[htbp]
\setlength\extrarowheight{2pt}
  \centering\footnotesize
  \caption{Properties of the  original  and the  BCL graphs.}
  \label{tab:meta}
  \begin{tabular}{|l|S|S|S|S|S|S[table-format=1.2e+2]|}%
    \hline
    \multicolumn{1}{|c|}{Graph} & 
    \multicolumn{2}{c|}{Size} & 
    \multicolumn{1}{c|}{\ Edges\ } & 
    \multicolumn{1}{c|}{\ Cats.\ } & 
    \multicolumn{1}{c|}{\ Buts.\ } & 
    \multicolumn{1}{c|}{\ Meta. \ } 
    \\
    \multicolumn{1}{|c|}{} & 
    \multicolumn{1}{c|}{$n^\lt$} & 
    \multicolumn{1}{c|}{$n^\rt$} & 
    \multicolumn{1}{c|}{$m$} & 
    \multicolumn{1}{c|}{$\ncat$} & 
    \multicolumn{1}{c|}{$\nbtf$} & 
    \multicolumn{1}{c|}{$c$}    
    \\ \hline
CondMat-Orig & 1.67e+04 & 2.20e+04 & 5.86e+04 & 1.24e+06& 7.05e+04 & 2.28e-01 \\ 
CondMat-CL & 1.67e+04 & 2.20e+04 & 5.86e+04 & 2.22e+06& 3.57e+02 & 6.43e-04 \\ 
\hline
IMDB-Orig & 1.28e+05 & 3.84e+05 & 1.47e+06 & 8.56e+08& 3.50e+06 & 1.64e-02 \\ 
IMDB-CL & 1.28e+05 & 3.84e+05 & 1.47e+06 & 1.11e+09& 1.41e+05 & 5.10e-04 \\ 
\hline
Flickr-Orig & 1.73e+06 & 1.04e+05 & 8.55e+06 & 2.57e+12& 3.53e+10 & 5.49e-02 \\ 
Flickr-CL & 1.73e+06 & 1.04e+05 & 8.39e+06 & 2.20e+12& 1.52e+10 & 2.78e-02 \\ 
\hline
MovieLens-Orig & 6.51e+04 & 7.16e+04 & 1.00e+07 & 2.46e+13& 1.20e+12 & 1.95e-01 \\ 
MovieLens-CL & 6.51e+04 & 7.16e+04 & 8.78e+06 & 1.37e+13& 5.34e+11 & 1.56e-01 \\ 
\hline
MillionSong-Orig & 1.02e+06 & 3.85e+05 & 4.84e+07 & 2.21e+13& 2.15e+11 & 3.89e-02 \\ 
MillionSong-CL & 1.02e+06 & 3.85e+05 & 4.81e+07 & 2.59e+13& 6.74e+10 & 1.04e-02 \\ 
\hline
Peer2Peer-Orig & 1.99e+06 & 5.38e+06 & 5.58e+07 & 8.18e+11& 3.80e+09 & 1.86e-02 \\ 
Peer2Peer-CL & 1.99e+06 & 5.38e+06 & 5.58e+07 & 1.20e+12& 1.14e+08 & 3.79e-04 \\ 
\hline
LiveJournal-Orig & 5.28e+06 & 7.49e+06 & 1.12e+08 & 3.36e+14& 3.30e+12 & 3.92e-02 \\ 
LiveJournal-CL & 5.28e+06 & 7.49e+06 & 1.11e+08 & 3.31e+14& 2.04e+12 & 2.47e-02 \\ 
\hline
  \end{tabular}
\end{table}

\section{Bipartite BTER Model}
\label{sec:bipartite-bter-model}

In the BTER model for one-way graphs, the goal is to match both the
degree distribution as well as the degreewise clustering coefficients
\cite{SeKoPi12,KoPiPlSe14}.
Our goal here is to extend those notions to the bipartite case.
We use \cref{eq:meta} as the bipartite definition of the
clustering coefficient, though other metrics exist as surveyed by
Latapy, Magnien, and Vecchio \cite{LaMaVe08} and Opsahl \cite{Op13}.

\subsection{Degreewise Metamorphosis Coefficient}
\label{sec:degr-metam-coeff}

BTER matches the degreewise clustering coefficient, so we need a
similar measure for bipartite graphs.  We describe the degreewise
metamorphosis coefficient, 
which provides a more nuanced measurement of bipartite community
structure than the metamorphosis coefficient. 
To the best of our knowledge, this idea has
not been proposed before.

We define the metamorphosis of an edge $(i,j)$ as
\begin{equation}\label{eq:cijdef} 
  c_{(i,j)} =
  \begin{cases}
    \displaystyle\frac{\nbtf_{(i,j)}}{\ncat_{(i,j)}}
    = \displaystyle\frac{\text{number of butterflies containing $(i,j)$}}
    {\text{number of caterpillars centered at $(i,j)$}}
    & \text{if }\ncat_{(i,j)} > 0, \\
    0 & \text{if } \ncat_{(i,j)}=0.
  \end{cases}
\end{equation}
We know the number of caterpillars centered at $(i,j)$ immediately from the degrees of its endpoints, i.e., 
\begin{equation}\label{eq:ncatdef}
  \ncat_{(i,j)} = (d_i^{\lt} -1) (d_j^{\rt} - 1).
\end{equation}
From this, we define the metamorphosis coefficients of vertices $i\in
U$ and $j\in V$ as the mean value over all edges incident to the
vertex:
\begin{equation}\label{eq:cidef}
  c^{\lt}_i = \frac{1}{d_i^{\lt}} \sum_{(i,j)\in E} c_{(i,j)}
  \qtext{and}
  c^{\rt}_j = \frac{1}{d_j^{\rt}} \sum_{(i,j)\in E} c_{(i,j)}.
\end{equation}
We may consider \cref{eq:cidef} to be the bipartite analogue of the clustering coefficient of a vertex, as introduced by Watts and Strogatz \cite{WaSt98}.
Finally, we can define the per-degree metamorphosis coefficients to be
\begin{equation}\label{eq:cddef} 
  c^{\lt}_d =   \frac{1}{|U_d|} \sum_{i \in U_d} c_i^{\lt}
  \qtext{and}
  c^{\rt}_d =   \frac{1}{|V_d|} \sum_{j \in V_d} c_j^{\rt},
\end{equation}
where $U_d$ and $V_d$ denote the subsets of degree-$d$ nodes, i.e.,
\begin{displaymath}
  U_d = \set{i \in U |d_i^{\lt} = d}
  \qtext{and}
  V_d = \set{j \in V |d_j^{\rt} = d}.
\end{displaymath}

Degreewise metamorphosis coefficients control for the effects of both
vertex mode and vertex degree on bipartite clustering. Accordingly,
this metric may reveal insights otherwise lost in metrics based only
on the ratio of total butterfly to caterpillar counts. To illustrate, 
consider the CondMat author-paper network, whose degreewise
metamorphosis coefficients are shown by the blue line in
\cref{fig:meta-CondMat-1,fig:meta-CondMat-2}.
The degreewise metamorphosis coefficients are also binned in the same
way that we binned the data for the degree distributions: We use
powers of two for the bins, so the $x$-coordinate $2^k$ corresponds to
the bin from $[2^k,2^{k+1})$. The $y$-coordinate is the average value
for that bin.  For the binning, we define $c_d^u = 0$ for any degree
such that $|U_d|=0$ (i.e., when there are no nodes of degree $d$), and
likewise for $c_c^v$.  
We see that degreewise metamorphosis coefficients in the author mode (\cref{fig:meta-CondMat-1})
are higher for low degrees than for high degrees, meaning that authors
with fewer papers tended to have higher proportion of repeat collaborations than
authors with many papers.
Conversely, the paper mode (\cref{fig:meta-CondMat-2}) shows that
authors of papers with few authors tend to have more repeats of the
same author set.

\subsection{Affinity Blocks}
\label{sec:affinity-blocks}

In BTER for one-way graphs, dense ER subgraphs are key
to producing triangles. For bipartite BTER, we will use dense bipartite ER
subgraphs to produce butterflies. We refer to these dense ER subgraphs as \emph{affinity blocks}.

In (one-way) BTER, an affinity block ideally consists of a set of
$d+1$ vertices of degree $d$. The connectivity of each block is computed 
according to the degree-$d$ clustering coefficient. The bipartite
affinity blocks are similar in spirit, but the bipartite nature of the
graph raises issues that require an entirely new approach to the block
construction.

The first key difference is that each affinity block in bipartite BTER
consists of {\em two} sets of vertices, one from each partition.
While each partition set in an affinity block ideally contains
vertices of the same degree, the degrees do not necessarily match
between partition sets. Indeed, in many bipartite graphs, one
partition set may have a very different range of degrees than the
other, so attempting to create blocks that match inter-partition
degree is not a realistic goal.

Consequently, a second key difference concerns how we determine the
{\em sizes} of the partition sets for the affinity blocks. In the
one-way BTER method, the size of each block only depends on the degree
of the vertices in the block. In the two-way case, the sizes of each
partition set within a block is more complicated.

\newcommand{\Gaf}{\hat G}
\newcommand{\Uaf}{\hat U}
\newcommand{\Vaf}{\hat V}
\newcommand{\Eaf}{\hat E}
\newcommand{\NUaf}{\hat n^{\lt}}
\newcommand{\NVaf}{\hat n^{\rt}}
\newcommand{\CUaf}{\hat c^{\lt}}
\newcommand{\CVaf}{\hat c^{\rt}}
\newcommand{\DUaf}{\hat d^{\lt}}
\newcommand{\DVaf}{\hat d^{\rt}}

To work out the calculations of sizes and connectivity for the blocks,
we consider building a single affinity block denoted by
$\Gaf = (\Uaf, \Vaf, \Eaf)$.
Without loss of generality, we assume all nodes in $\Uaf$ want
degree $\DUaf$ and all nodes in $\Vaf$ want degree
$\DVaf$. Note that these degrees are with respect to the entire graph,
not the subgraph.
We further assume all nodes in $\Uaf$ want metamorphosis coefficient
$\CUaf$ and likewise for $\Vaf$ and $\CVaf$. When matching a
real-world graph, we choose the degreewise metamorphosis coefficients
corresponding to the target
degrees as defined in \cref{eq:cddef}, i.e.,
\begin{displaymath}
  \CUaf = c_{\DUaf} \qtext{and} 
  \CVaf = c_{\DVaf}.
\end{displaymath}
The goal is to determine the sizes
$\NUaf= | \Uaf |$ and $\NVaf= | \Vaf |$ and the connectivity, $\rho$, which is the probability of an edge  between a node in $\NUaf$ and a vertex in $\NVaf$, and thus  the number of edges, $| \Eaf |$

For $i \in \Uaf$, we can compute its expected metamorphosis
coefficient as follows. By definition \cref{eq:cidef}, we have
\begin{equation}\label{eq:c1}
  c_i^\lt  = \frac{1}{d_i^\lt} \sum_{(i,j)\in E} c_{(i,j)} 
  = \frac{1}{d_i^\lt} \left( \sum_{(i,j) \in \Eaf}  c_{(i,j)}  + \sum_{(i,j) \in E \setminus \Eaf}  c_{(i,j)} \right)
 \approx  \frac{1}{d_i^\lt} \sum_{(i,j) \in \Eaf}  c_{(i,j)} .
\end{equation}
The last step comes from the assumption that nearly all butterflies in the larger graph come from the affinity blocks.
Using definition \cref{eq:cijdef}, we can rewrite \cref{eq:c1} as
\begin{equation}
  \label{eq:c2}
  c_i^\lt 
  = \frac{1}{d_i^\lt} \sum_{(i,j) \in \Eaf}  \frac{\nbtf_{(i,j)}}{\ncat_{(i,j)}} 
  = \frac{1}{d_i^\lt} \sum_{(i,j) \in \Eaf} \frac{\nbtf_{(i,j)}}{(d_i^\lt-1)(d_j^\rt-1)}.
\end{equation}
We have assumed that the degrees and clustering coefficients within
$\Gaf$ are constant, so \cref{eq:c2} becomes
\begin{equation}
  \label{eq:c3}
  \CUaf = \frac{1}{\DUaf(\DUaf-1)(\DVaf-1)} \sum_{(i,j)\in \Eaf} \nbtf_{(i,j)} 
  = \frac{2\nbtf_i}{\DUaf(\DUaf-1)(\DVaf-1)} .
\end{equation}
The last equality uses the fact that the sum of butterflies involving
edges of the form $(i,j)$ is equal to two times the number of
butterflies involving node $i$ since each such butterfly has two edges
involving $i$.
In expectation, 
\begin{equation}\label{eq:c4} 
  \nbtf_i \doteq \rho^4 (\NUaf - 1) {\NVaf \choose 2},
\end{equation}
because there are $(\NUaf - 1)$ other choices for the second node in
$\Uaf$ and ${\NVaf \choose 2}$ choices for the two nodes in $\Vaf$.
Finally, $\rho^4$ is the probability that all four edges exist.
Combining \cref{eq:c3,eq:c4} gives
\begin{equation}
  \label{eq:c5}
  \CUaf \doteq \frac{\rho^4 \NVaf (\NUaf-1)(\NVaf-1)}{\DUaf(\DUaf-1)(\DVaf-1)}.
\end{equation}
Using analogous reasoning for $\CVaf$, we ultimately want
\begin{equation}\label{eq:rho4}
  \rho^4 
  = \frac{\CUaf \DUaf (\DUaf-1)(\DVaf-1)}{\NVaf(\NUaf-1)(\NVaf-1)}
  = \frac{\CVaf \DVaf (\DUaf-1)(\DVaf-1)}{\NUaf(\NUaf-1)(\NVaf-1)}
\end{equation}
Ideally, we would have
\begin{equation}\label{eq:neq}
  \NUaf = \DVaf
  \qtext{and}
  \NVaf = \DUaf,
\end{equation}
but we cannot generally satisfy \cref{eq:rho4,eq:neq} at the same time. Therefore, we
choose one of the equalities in \cref{eq:neq} and then solve for the
other number of nodes and connectivity using \cref{eq:rho4} to get
\begin{align}
  \label{eq:opta}
  \NUaf = \DVaf & \quad\Rightarrow\quad \NVaf = \frac{\CUaf}{\CVaf} \DUaf, 
  \quad \rho = \frac{(\DUaf-1)(\CVaf)^2}{\CUaf\DUaf-\CVaf}
    \\
  \label{eq:optb}
  \NVaf = \DUaf & \quad\Rightarrow\quad \NUaf = \frac{\CVaf}{\CUaf} \DVaf.
  \quad \rho = \frac{(\DVaf-1)(\CUaf)^2}{\CVaf\DVaf-\CUaf}.
\end{align}

So that we can have the possibility of using a complete bipartite
subgraph to yield metamorphosis coefficients of one, we constrain our
choices to satisfy:
\begin{equation}\label{eq:ni}
  \NUaf \geq \DVaf
  \qtext{and}
  \NVaf \geq \DUaf.
\end{equation}
To satisfy \cref{eq:ni}, we choose \cref{eq:opta} if $\frac{\CUaf}{\CVaf}\geq 1$ and \cref{eq:optb} otherwise.
It is easy to see from \cref{eq:rho4} that this has the added bonus of
ensuring $\rho \leq 1$.
This logic forms the basis of building affinity blocks for bipartite BTER. 

\subsection{Bipartite BTER algorithm}
\label{sec:bipart-bter-algor}

As explained in \cref{sec:affinity-blocks}, the affinity block building process for bipartite BTER takes into account both the desired vertex degrees and 
corresponding desired per-degree metamorphosis coefficients when setting the partition sizes and connectivity of each affinity block. 
This information is then used to model each affinity block as an ER subgraph, where the preponderance of the graph butterflies are created.
Accordingly, the affinity block construction is key to matching per-degree metamorphosis coefficients. In order to match the desired degree distribution,
the remaining \emph{excess degree} (i.e., edges not used in constructing the affinity blocks) is connected via a fast bipartite CL procedure. The full 
bipartite BTER algorithm is listed in \cref{alg:bibter}.

\begin{algorithm}[htbp]
  \caption{Bipartite BTER}
  \label{alg:bibter}
  \begin{algorithmic}[1]
    \Procedure{bibter}{$\set{ d^{\lt}_i }$, $\set{ d_j^{\rt} }$, $\set{ c_d^\lt}$, $\set{ c_d^\rt}$ }
    \Statex \emph{We assume that degrees are sorted in increasing order}
    \State $m \gets \sum_i d_i^{\lt}$
    \State $E \gets \emptyset$
    \State $\set{ e^{\lt}_i } \gets \set{ d^{\lt}_i }$, $\set{ e_j^{\rt} } \gets \set{ d_j^{\rt} }$ \Comment{Excess degree initialization}
    \State $i \gets \min \set{i | d_i^\lt > 1 }$,
    $j \gets \min \set{j | d_j^\rt > 1 }$
    \Repeat \Comment{Create affinity blocks until nodes exhausted}
    \State $\DUaf \gets d_i^\lt$, $\DVaf \gets d_j^\rt$
    \State $\CUaf \gets c_{\DUaf}$, $\CVaf \gets c_{\DVaf}$
    \If{$\CUaf/\CVaf \geq 1$}
    \State $\NUaf \gets \DVaf$, $\NVaf \gets \texttt{round}\left(\frac{\CUaf}{\CVaf} \DUaf\right)$, 
    $\rho \gets \left(\frac{(\DUaf-1)(\CVaf)^2}{\CUaf\DUaf-\CVaf}\right)^{1/4}$
    \Else
    \State $\NVaf \gets \DUaf$, $\NUaf \gets \texttt{round}\left(\frac{\CVaf}{\CUaf} \DVaf\right)$,
    $\rho \gets \left(\frac{(\DVaf-1)(\CUaf)^2}{\CVaf\DVaf-\CUaf}\right)^{1/4}$
    \EndIf
    \If{$i+\NUaf-1 \leq n^\lt$ and $j+\NVaf-1 \leq n^\rt$} \Comment Create ER subgraph
    \For{$\hat i = i,i+1,\dots,i+\NUaf-1$}
    \For{$\hat j = j,j+1,\dots,j+\NVaf-1$}
    \State $r \gets U(0,1)$ \Comment Uniform random value in $[0,1]$
    \If{$r <= \rho$}
    \State $E \gets E \cup (\hat i, \hat j)$
    \State $e_{\hat i}^\lt = \max\{0,e_{\hat i}^\lt - 1\}$, $e_{\hat j}^\rt = \max\{0,e_{\hat j}^\rt - 1\}$ \Comment{Update excess degree}
    \EndIf
    \EndFor
    \EndFor
    \EndIf
    \State $i \gets i + \NUaf$, $j \gets j + \NVaf$ 
    \Until{$i > n^\lt$ or $j >  n^\rt$}
    \State $E \gets E \cup \textsc{fbcl}(\set{ e^{\lt}_i }, \set{ e_j^{\rt} })$
    \EndProcedure
  \end{algorithmic}
\end{algorithm}

\subsection{Experimental Results}
\label{sec:experimental-results}

\begin{figure}[htbp]
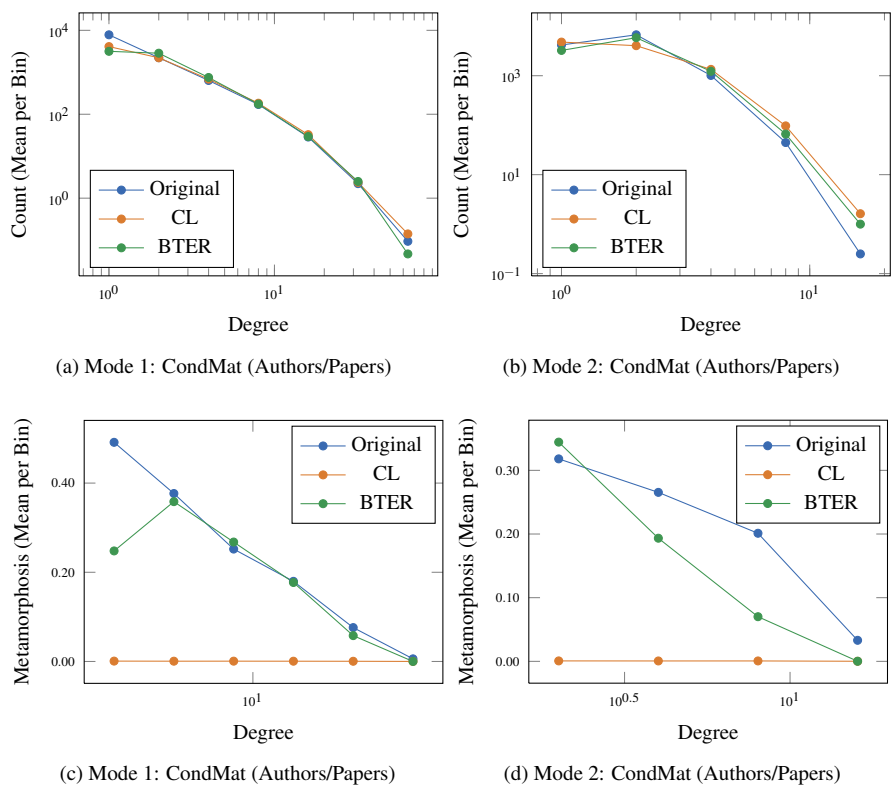

  \centering
  \DegDistSubFig*{CondMat}{Authors}{Papers}{1}(ddbter)
  \DegDistSubFig*{CondMat}{Authors}{Papers}{2}(ddbter)

  \MetaSubFig[mbin]*{CondMat}{Authors}{Papers}{1}
  \MetaSubFig[mbin]*{CondMat}{Authors}{Papers}{2}
  \caption{Degree distribution and degreewise metamorphosis
    coefficients on the original CondMat graph as well as the models
    generated by CL and bipartite BTER.}
  \label{fig:condmat}
\end{figure}

We generate bipartite BTER  graphs using the procedure described in
\cref{sec:bipart-bter-algor}.  We first discuss a single graph:
CondMat. We show the resulting degree distribution and degreewise
metamorphosis coefficients in \cref{fig:condmat}.  
The degree distributions, shown in
\cref{fig:ddbter-CondMat-1,fig:ddbter-CondMat-2}, show little
difference between bipartite BTER and CL; both are good matches to the original
degree distribution.
The degreewise metamorphosis coefficients are shown in
\cref{fig:meta-CondMat-1,fig:meta-CondMat-2}. Although there is not a
perfect match between bipartite BTER and the original graph, it is much better
than CL, which has almost no butterflies.

Summary data for all graphs is shown in \cref{tab:metabter}. This is the same as \cref{tab:meta} except that now we have added a row for bipartite BTER. 
For the first four graphs, we are reporting average values over 100 experiments, and we also report the entire range of values;
we do not do multiple trials for the larger graphs because the postprocessing to count the butterflies is extremely expensive.
The number of butterflies and metamorphosis coefficients are significantly improved as compared to CL. We see that CondMat has 70,000 butterflies, bipartite CL produces less than 400 butterflies, but bipartite BTER produces 120,000 butterflies. The bipartite BTER number is a slight overestimate, but overall much better than CL.

\sisetup{scientific-notation = true,group-separator={},table-format=1.2e+1,output-exponent-marker = \text{e},exponent-product={}}%
\begin{table}[thbp]
\setlength\extrarowheight{2pt}
  \centering\footnotesize
  \caption{Metamorphosis coefficients for original versus BCL graphs.}
  \label{tab:metabter}
  \begin{tabular}{|l|S|S|S|S|S|S[table-format=1.2e+2]|}%
    \hline
    \multicolumn{1}{|c|}{Graph} & 
    \multicolumn{2}{c|}{Size} & 
    \multicolumn{1}{c|}{Edges} & 
    \multicolumn{1}{c|}{Cats.} & 
    \multicolumn{1}{c|}{Buts.} & 
    \multicolumn{1}{c|}{Meta.} 
    \\
    \multicolumn{1}{|c|}{} & 
    \multicolumn{1}{c|}{$n^\lt$} & 
    \multicolumn{1}{c|}{$n^\rt$} & 
    \multicolumn{1}{c|}{$m$} & 
    \multicolumn{1}{c|}{$\ncat$} & 
    \multicolumn{1}{c|}{$\nbtf$} & 
    \multicolumn{1}{c|}{$c$}    
    \\ \hline
CondMat-Orig & 1.67e+04 & 2.20e+04 & 5.86e+04 & 1.24e+06& 7.05e+04 & 2.28e-01 \\ 
CondMat-CL & 1.67e+04 & 2.20e+04 & 5.86e+04 & 2.22e+06& 3.57e+02 & 6.43e-04 \\ 
CondMat-BTER & 1.67e+04 & 2.20e+04 & 6.00e+04 & 2.36e+06& 1.10e+05 & 1.86e-01 \\ 
\multicolumn{1}{|r|}{\scriptsize BTER 100 Trial Range} & \multicolumn{1}{@{\,}c@{\,}|}{\scriptsize[1.67\,--\,1.67]} & \multicolumn{1}{@{\,}c@{\,}|}{\scriptsize[2.20\,--\,2.20]} & \multicolumn{1}{@{\,}c@{\,}|}{\scriptsize[6.00\,--\,6.01]} & \multicolumn{1}{@{\,}c@{\,}|}{\scriptsize[2.31\,--\,2.42]} & \multicolumn{1}{@{\,}c@{\,}|}{\scriptsize[1.01\,--\,1.16]} & \multicolumn{1}{@{\,}c@{\,}|}{\scriptsize[1.73\,--\,1.99]} \\ 
\hline
IMDB-Orig & 1.28e+05 & 3.84e+05 & 1.47e+06 & 8.56e+08& 3.50e+06 & 1.64e-02 \\ 
IMDB-CL & 1.28e+05 & 3.84e+05 & 1.47e+06 & 1.11e+09& 1.41e+05 & 5.10e-04 \\ 
IMDB-BTER & 1.28e+05 & 3.84e+05 & 1.47e+06 & 1.35e+09& 6.77e+06 & 2.00e-02 \\ 
\multicolumn{1}{|r|}{\scriptsize BTER 100 Trial Range} & \multicolumn{1}{@{\,}c@{\,}|}{\scriptsize[1.28\,--\,1.28]} & \multicolumn{1}{@{\,}c@{\,}|}{\scriptsize[3.84\,--\,3.84]} & \multicolumn{1}{@{\,}c@{\,}|}{\scriptsize[1.47\,--\,1.47]} & \multicolumn{1}{@{\,}c@{\,}|}{\scriptsize[1.35\,--\,1.36]} & \multicolumn{1}{@{\,}c@{\,}|}{\scriptsize[6.62\,--\,6.91]} & \multicolumn{1}{@{\,}c@{\,}|}{\scriptsize[1.95\,--\,2.04]} \\ 
\hline
Flickr-Orig & 1.73e+06 & 1.04e+05 & 8.55e+06 & 2.57e+12& 3.53e+10 & 5.49e-02 \\ 
Flickr-CL & 1.73e+06 & 1.04e+05 & 8.39e+06 & 2.20e+12& 1.52e+10 & 2.78e-02 \\ 
Flickr-BTER & 3.96e+05 & 1.04e+05 & 8.34e+06 & 2.36e+12& 4.25e+10 & 7.21e-02 \\ 
\multicolumn{1}{|r|}{\scriptsize BTER 100 Trial Range} & \multicolumn{1}{@{\,}c@{\,}|}{\scriptsize[3.96\,--\,3.96]} & \multicolumn{1}{@{\,}c@{\,}|}{\scriptsize[1.04\,--\,1.04]} & \multicolumn{1}{@{\,}c@{\,}|}{\scriptsize[8.34\,--\,8.34]} & \multicolumn{1}{@{\,}c@{\,}|}{\scriptsize[2.35\,--\,2.37]} & \multicolumn{1}{@{\,}c@{\,}|}{\scriptsize[4.23\,--\,4.27]} & \multicolumn{1}{@{\,}c@{\,}|}{\scriptsize[7.19\,--\,7.23]} \\ 
\hline
MovieLens-Orig & 6.51e+04 & 7.16e+04 & 1.00e+07 & 2.46e+13& 1.20e+12 & 1.95e-01 \\ 
MovieLens-CL & 6.51e+04 & 7.16e+04 & 8.78e+06 & 1.37e+13& 5.34e+11 & 1.56e-01 \\ 
MovieLens-BTER & 1.07e+04 & 6.99e+04 & 8.52e+06 & 1.23e+13& 1.08e+12 & 3.51e-01 \\ 
\multicolumn{1}{|r|}{\scriptsize BTER 100 Trial Range} & \multicolumn{1}{@{\,}c@{\,}|}{\scriptsize[1.07\,--\,1.07]} & \multicolumn{1}{@{\,}c@{\,}|}{\scriptsize[6.99\,--\,6.99]} & \multicolumn{1}{@{\,}c@{\,}|}{\scriptsize[8.51\,--\,8.52]} & \multicolumn{1}{@{\,}c@{\,}|}{\scriptsize[1.23\,--\,1.23]} & \multicolumn{1}{@{\,}c@{\,}|}{\scriptsize[1.07\,--\,1.08]} & \multicolumn{1}{@{\,}c@{\,}|}{\scriptsize[3.50\,--\,3.51]} \\ 
\hline
MillionSong-Orig & 1.02e+06 & 3.85e+05 & 4.84e+07 & 2.21e+13& 2.15e+11 & 3.89e-02 \\ 
MillionSong-CL & 1.02e+06 & 3.85e+05 & 4.81e+07 & 2.59e+13& 6.74e+10 & 1.04e-02 \\ 
MillionSong-BTER & 1.02e+06 & 3.85e+05 & 4.80e+07 & 2.93e+13& 1.90e+11 & 2.59e-02 \\ 
\hline
Peer2Peer-Orig & 1.99e+06 & 5.38e+06 & 5.58e+07 & 8.18e+11& 3.80e+09 & 1.86e-02 \\ 
Peer2Peer-CL & 1.99e+06 & 5.38e+06 & 5.58e+07 & 1.20e+12& 1.14e+08 & 3.79e-04 \\ 
Peer2Peer-BTER & 1.99e+06 & 5.38e+06 & 5.60e+07 & 1.66e+12& 6.71e+09 & 1.61e-02 \\ 
\hline
LiveJournal-Orig & 5.28e+06 & 7.49e+06 & 1.12e+08 & 3.36e+14& 3.30e+12 & 3.92e-02 \\ 
LiveJournal-CL & 5.28e+06 & 7.49e+06 & 1.11e+08 & 3.31e+14& 2.04e+12 & 2.47e-02 \\ 
LiveJournal-BTER & 3.20e+06 & 7.49e+06 & 1.10e+08 & 3.50e+14& 4.61e+12 & 5.26e-02 \\ 
\hline
  \end{tabular}
\end{table}

The degreewise metamorphosis coefficients for the remaining graphs are
shown in \cref{fig:meta1,fig:meta2}.  Even in cases where bipartite CL's overall
numbers are close to the original graph as shown in
\cref{tab:metabter}, the degreewise metamorphosis coefficients are
essentially zero. Bipartite BTER corrects this problem and gets metamorphosis coefficients
that are much closer to what we see in the original graph.  In some cases
like MovieLens (\cref{fig:meta-MovieLens-1}), the CL metamorphosis
coefficients are much higher than zero, which is the result of the overall high
density of the graph. Indeed, it has been proven (see Lemma 1 in \cite{TaVe16}) that under mild assumptions, \emph{any} bipartite graph with $m$ edges and partition sizes $n^u$ and $n^v$ contains at least on the order of $(\frac{n^u}{n^v})^2\cdot (\frac{m}{n^u+n^v})^4$ many butterflies. Thus, the presence of a certain minimum number butterflies in bipartite graphs of sufficient density is inevitable; nevertheless, the original graph still has higher values that are matched better by bipartite BTER.  MovieLens also has an
unusual profile in Mode 2 (see \cref{fig:meta-MovieLens-2}), which is really just an artifact of the
data collection. Nevertheless, bipartite BTER is able to obtain a reasonable
approximation. 
For completeness, the degree distributions for bipartite CL and BTER as compared
to the original graphs are shown in \cref{fig:ddbter1,fig:ddbter2}.
There is little difference between bipartite BTER and CL in terms of the degree distribution.

\newcommand\metacaption{Degreewise metamorphosis coefficient of the original graph (\NameColorA), fast BCL (\NameColorB), and bipartite BTER (\NameColorC).}

\begin{figure}[phtb]
  \centering 
  \MetaSubFig[mbin]*{IMDB}{Actors}{Movies}{1}
  \MetaSubFig[mbin]*{IMDB}{Actors}{Movies}{2}

  \MetaSubFig[mbin]*{Flickr}{Users}{Groups}{1}
  \MetaSubFig[mbin]*{Flickr}{Users}{Groups}{2}

  \MetaSubFig[mbin]*{MovieLens}{Movies}{Critics}{1}
  \MetaSubFig[mbin]*{MovieLens}{Movies}{Critics}{2}
  \caption{\metacaption}
  \label{fig:meta1}
\end{figure}
\begin{figure}[phtb]
  \centering
  \MetaSubFig[mbin]*{MillionSong}{Users}{Songs}{1}
  \MetaSubFig[mbin]*{MillionSong}{Users}{Songs}{2}

  \MetaSubFig[mbin]*{Peer2Peer}{Peers}{Files}{1}
  \MetaSubFig[mbin]*{Peer2Peer}{Peers}{Files}{2}

  \MetaSubFig[mbin]*{LiveJournal}{Users}{Groups}{1}
  \MetaSubFig[mbin]*{LiveJournal}{Users}{Groups}{2}
  \caption{\metacaption}
  \label{fig:meta2}
\end{figure}
\newcommand{\ddbtercaption}{Degree distributions illustrating the original (\NameColorA), fast BCL (\NameColorB), and bipartite BTER (\NameColorC). The data is log-binned.}%
\begin{figure}[phtb]
  \centering
  \DegDistSubFig*{IMDB}{Actors}{Movies}{1}(ddbter)
  \DegDistSubFig*{IMDB}{Actors}{Movies}{2}(ddbter)

  \DegDistSubFig*{Flickr}{Users}{Groups}{1}(ddbter)
  \DegDistSubFig*{Flickr}{Users}{Groups}{2}(ddbter)

  \DegDistSubFig*{MovieLens}{Movies}{Critics}{1}(ddbter)
  \DegDistSubFig*{MovieLens}{Movies}{Critics}{2}(ddbter)
  \caption{\ddbtercaption}
  \label{fig:ddbter1}
\end{figure}
\begin{figure}[phtb]
  \centering
  \DegDistSubFig*{MillionSong}{Users}{Songs}{1}(ddbter)
  \DegDistSubFig*{MillionSong}{Users}{Songs}{2}(ddbter)

  \DegDistSubFig*{Peer2Peer}{Peers}{Files}{1}(ddbter)
  \DegDistSubFig*{Peer2Peer}{Peers}{Files}{2}(ddbter)
 
  \DegDistSubFig*{LiveJournal}{Users}{Groups}{1}(ddbter)
  \DegDistSubFig*{LiveJournal}{Users}{Groups}{2}(ddbter)
   \caption{\ddbtercaption}
  \label{fig:ddbter2}
\end{figure}

\section{Conclusions}
\label{sec:conclusions}

We have considered the problem of how to generate realistic bipartite
graphs to reproduce the characteristics of large real-world
networks. Our first model, bipartite CL, accurately
reproduces the degree distribution. Our second model,
bipartite BTER, goes further to capture the degreewise metamorphosis
coefficients. High coefficients are indicative of a relatively large
number of butterflies indicating cohesion or community structure,
which is rare in sparse random graphs unless there are some behaviors that go
beyond just the degree structure. Creating realistic graph models
leads to some hypotheses about the ways the graphs were formed. In the
cases where CL greatly underestimated the number of butterflies
(CondMat, IMDB, and Peer2Peer), we can surmise that there is some
significant community-like behavior. This is easy to see for
authorship of papers (CondMat) and actors appearing in movies
(IMBD). For the Peer2Peer network, we might hypothesize that some
tight-knit peer groups are sharing many files between themselves. The
other graphs have a smaller difference between the number of
butterflies produced by CL and the real-world graph, indicating that
there is much less community structure. In fact, some graphs are so
dense that CL has a nonzero metamorphosis coefficient, especially
MovieLens.

Although these models match real-world observations in some aspects,
more study is needed to understand their limitations. 
How well do these models reproduce other metrics such as graph diameter, singular values of the adjacency matrix, centrality measures, 
joint degree distributions, assortativity, subgraph census,
and so on?
Can we find
evidence that affinity blocks exist in real-world graphs? We have
created non-overlapping blocks; is that realistic? 
We have not
mentioned time-evolution, but certainly all networks are evolving in
time and we need models that capture such changes. These are but a few
topics for future investigation. 

\appendix

\section*{Funding}
This work was supported in part by the Defense Advanced Research Project Agency (DARPA).
Sandia National Laboratories is a multi-program laboratory
managed and operated by Sandia Corporation, a wholly owned
subsidiary of Lockheed Martin Corporation, for the U.S. Department
of Energy's National Nuclear Security Administration under
contract DE-AC04-94AL85000.%

\section*{Acknowledgments}
We thank our colleagues for helpful discussions in the course of this
work, especially Grey Ballard, Kenny Chowdhary, Danny Dunlavy, Kevin
Matulef, and Michael Wolf.
We also thank the two anonymous referees for insightful comments that
improved the work considerably.

\bibliographystyle{comnet-mod}

\end{document}